\documentclass[twocolumn]{aastex62}

\usepackage{gensymb}
\usepackage[version=4]{mhchem}
\shorttitle{NUV secondary eclipse observation of KELT-9b}
\shortauthors{Hooton et al.}

\begin{document}

\title{A GROUND-BASED NUV SECONDARY ECLIPSE OBSERVATION OF KELT-9B}

\correspondingauthor{Matthew Hooton}
\email{mhooton01@qub.ac.uk}

\author[0000-0003-0030-332X]{Matthew J. Hooton}
\affil{Astrophysics Research Centre, 
School of Mathematics and Physics,
Queen's University Belfast,
Belfast BT7 1NN, UK}

\author{Christopher A. Watson}
\affil{Astrophysics Research Centre, 
School of Mathematics and Physics,
Queen's University Belfast,
Belfast BT7 1NN, UK}

\author[0000-0001-6391-9266]{Ernst J. W. de Mooij}
\affil{School of Physical Sciences, 
Dublin City University,
Glasnevin,
Dublin 9, Ireland}

\author[0000-0002-9308-2353]{Neale P. Gibson}
\affil{Astrophysics Research Centre, 
School of Mathematics and Physics,
Queen's University Belfast,
Belfast BT7 1NN, UK}

\author[0000-0003-4269-3311]{Daniel Kitzmann}
\affil{University of Bern,
Center for Space and Habitability,
Gesellschaftsstrasse 6,
CH-3012, Bern, Switzerland.}

\begin{abstract}

KELT-9b is a recently discovered exoplanet with a 1.49 d orbit around a B9.5/A0-type star. The unparalleled levels of UV irradiation it receives from its host star put KELT-9b in its own unique class of ultra-hot Jupiters, with an equilibrium temperature $>$ 4000 K. The high quantities of dissociated hydrogen and atomic metals present in the dayside atmosphere of KELT-9b bear more resemblance to a K-type star than a gas giant. We present a single observation of KELT-9b during its secondary eclipse, taken with the Wide Field Camera on the Isaac Newton Telescope (INT). This observation was taken in the U-band, a window particularly sensitive to Rayleigh scattering. We do not detect a secondary eclipse signal, but our 3$\sigma$ upper limit of 181 ppm on the depth allows us to constrain the dayside temperature of KELT-9b at pressures of $\sim$30 mbar to 4995 K (3$\sigma$). Although we can place an observational constraint of $A_g<$ 0.14, our models suggest that the actual value is considerably lower than this due to \ce{H-} opacity. This places KELT-9b squarely in the albedo regime populated by its cooler cousins, almost all of which reflect very small components of the light incident on their daysides. This work demonstrates the ability of ground-based 2m-class telescopes like the INT to perform secondary eclipse studies in the NUV, which have previously only been conducted from space-based facilities.

\end{abstract}

\keywords{planets and satellites: atmospheres --- stars: individual (KELT-9) --- techniques: photometric --- ultraviolet: planetary systems }

\section{Introduction} \label{sec:intro}

The measurement of the drop in flux of an exoplanet-star pair when the planet is occulted by its host star has established itself as an important tool to study the atmospheres of exoplanets. At near-infrared wavelengths and longer, thermal emission is the dominant source of flux from hot Jupiters \citep{2007ApJ...667L.191L}. Measurements of thermal emission have led to the detection of atmospheric features such as global heat redistribution \citep{2007Natur.447..183K}, the presence of a temperature inversion \citep{2017Natur.548...58E} and atmospheric variability \citep{2016NatAs...1E...4A}. 

At optical wavelengths and shorter, the component of flux from hot Jupiters due to thermal emission drops off sharply, such that the dominant component of flux is expected to be due to light reflected from its host. Measurements of thermal emission for various hot Jupiters imply that they should have reflection signatures sufficiently large to be detectable with current instrumentation \citep[e.g.][]{2015MNRAS.449.4192S,Schwartz:2017kt}. However, the vast majority of searches for reflected light from hot Jupiters at optical wavelengths---where their host stars typically emit most of their energy---have resulted in non-detections \citep[e.g.][]{CollierCameron:2002bi,2003MNRAS.344.1271L,2008ApJ...689.1345R,2013A&A...557A..74G,2017AJ....153...40D,2018AJ....156...44M}. These results are consistent with predictions that scattering in the optical is suppressed by alkali absorption for cloud-free atmospheres \citep{2000ApJ...538..885S,2008ApJ...678.1436B}. 

To date, two studies have utilised the capabilities of HST/STIS to observe secondary eclipses of hot Jupiters in the NUV. This wavelength range is potentially more favourable than the optical for detecting reflected light from exoplanets orbiting hot stars, as alkali absorption is much weaker and the Rayleigh scattering cross-section is much higher. Whilst \citet{2017ApJ...847L...2B} did not detect reflected light at NUV wavelengths for WASP-12b ($T_{eq}\sim$ 2,500 K), \citet{2013ApJ...772L..16E} measured a geometric albedo ($A_g$) of 0.40$\pm$0.12 at 290-450 nm (a wavelength range overlapping with the U-band) for HD 189733 b ($T_{eq}\sim$ 1,200 K). These results support studies \citep[e.g.][]{2016ApJ...826L..16H,2016ApJ...817L..16S,2017MNRAS.464.4247W} suggesting that the most highly irradiated planets are less likely to have clouds in their atmospheres, as well as observational evidence for clouds in the atmosphere of HD 189733 b \citep{Pont:2008ft,2011MNRAS.416.1443S}.

The 4050 K equilibrium temperature of the recently-discovered KELT-9b \citep{2017Natur.546..514G} is by far the hottest of any known exoplanet. Its 1.49 day orbit around HD 195689, a B9.5/A0-type star, means that KELT-9b is more heavily-irradiated at UV wavelengths than any other known exoplanet. \citet{Hoeijmakers:2018ir} obtained high-resolution spectra of KELT-9b during its transit and detected Fe, Fe\textsuperscript{+} and Ti\textsuperscript{+} features with high significance, suggesting a temperature in excess of 4000 K at the terminator. The 4600$\pm$150 K dayside temperature \citep{2017Natur.546..514G} measured from its z'-band eclipse depth (Collins et al. 2019, in prep.) is comparable to that of a K4-type star. This high dayside temperature means that KELT-9b is the only known planet expected to have a U-band eclipse depth $>$ 50 ppm due to thermal emission.

In this letter, we present a photometric ground-based U-band secondary eclipse of KELT-9b, which allows us to constrain the energy budget of this unique exoplanet. In chapter \ref{sec:observations} we summarise our observation, in chapter \ref{sec:reduction} we describe the steps taken to reduce the data, in chapter \ref{sec:results} we describe how we fitted the eclipse depth and in chapter \ref{sec:discussion} we summarise how we modelled the KELT-9b spectrum and discuss our result, along with the wider implications of this study.

\section{Observation}
\label{sec:observations}

We observed one secondary eclipse of KELT-9b on 2017 July 20 using the Wide Field Camera (WFC) with the U-band filter on the 2.5m Isaac Newton Telescope (INT) at the Observatorio del Roque de los Muchachos on the island of La Palma. The observations lasted $\sim$ 8.3 hours and started at 21:15:29 UT. During this time, 326 frames were obtained with an average cadence of 75.1 seconds and an exposure time of 45.1 seconds. 155 of these frames were taken when KELT-9b was fully or partially occulted by its host. The observations commenced and concluded in evening and morning twilights, respectively. 32 frames taken during twilight, when KELT-9b was out-of-eclipse, were removed as the increased sky brightness caused strong systematics in those sections of the light curves. For 17 frames that were randomly distributed through the night, we observed that the exposure time was only $\sim$44.6 seconds. However, due to our use of differential photometry, no visible correlation was observed between exposure time and flux once the target had been normalised with the comparison stars. Although the WFC consists of a 4 CCD mosaic --- each with a pixel scale of 0.33'' per pixel --- only the central CCD (CCD4) was used, giving us a field-of-view of $\sim$22.7' by 11.4'.

We performed the observations with the telescope defocused, which acts to reduce overheads, minimise errors associated with flat fielding and make the resulting PSFs less sensitive to variations in seeing. This resulted in a donut-shaped PSF with a diameter of 54 pixels (18''). Due to the defocusing, the telescope auto-guider was not used. Instead, a custom code that uses the science frames to account for telescope drift was used. Care was taken to ensure that the target and the most promising comparison stars were positioned on well-behaved parts of the detector and the drift throughout the night was less than 4 pixels (1.3''). 


\begin{figure*}[t]
\gridline{\fig{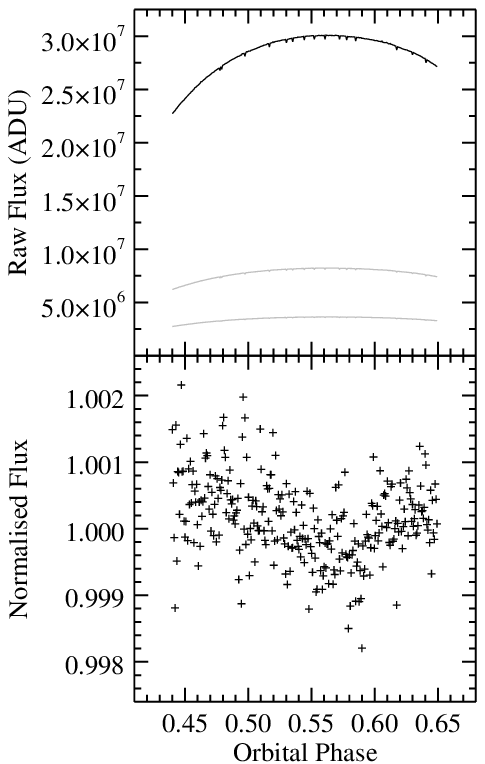}{0.3\textwidth}{(a)}
\fig{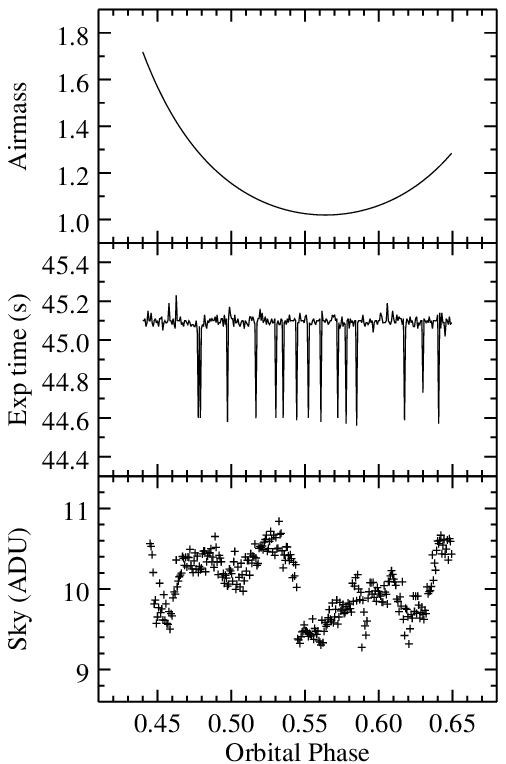}{0.3\textwidth}{(b)}
\fig{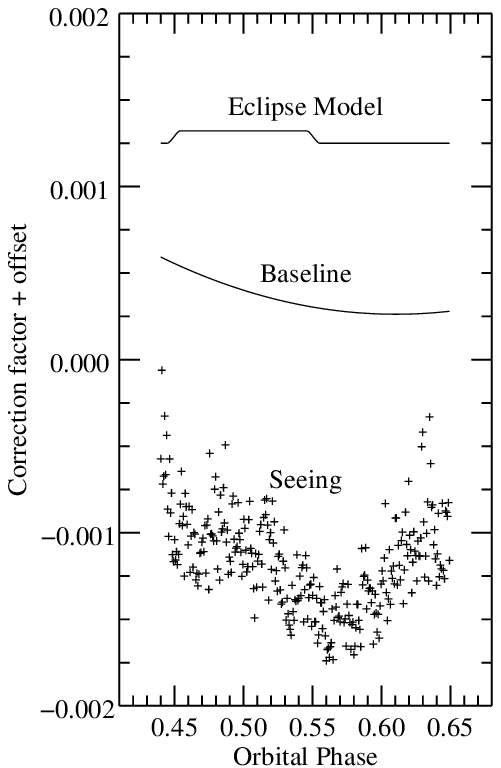}{0.3\textwidth}{(c)}
}
\caption{\textbf{(a)} Top panel: the raw light curves of KELT-9 (black points) and two comparison stars (grey points). Bottom panel: the normalised light curve prior to the MCMC fit. \textbf{(b)} A selection of parameters where no strong correlations were observed with the normalised light curve. \textbf{(c)} Components fitted to the normalised light curve using MCMC. }\label{Fig1}
\end{figure*}

\section{Data Reduction}
\label{sec:reduction}

Each of the images was overscan-subtracted on a row-by-row basis using the mean of the overscan regions at either side of the CCD. The row by row bias subtraction was used to correct for a known issue with the WFC that affected about half of the frames, in which the bias level present in the frames drops and corrects itself after a period of time. This was followed by a full-frame bias subtraction. The images were then each flat fielded using a master flat constructed from twilight flats. Finally, a second-order polynomial was fit to and subtracted from the entirety of each frame with the stars masked, to remove the small gradient in the sky background across the  CCD.

We observed crosstalk between the 4 CCDs that make up the WFC mosaic, which caused bright stars from one CCD to be ghosted onto the same position on other CCDs. This caused the addition or subtraction of $<$ 3 ADU on a background of $\sim$ 2,000 ADU in the raw frames. However, the target and comparison stars did not fall on any of the affected regions.  

Finally, we performed aperture photometry on the target and each of the two comparison stars using an aperture with a radius of 51 pixels, selected to maximise the flux and minimise the influence from the background. The annuli used to subtract the residual sky background from each star had inner and outer radii of 72 and 91 pixels, respectively. 

\section{Analysis \& Results}
\label{sec:results}

\begin{figure*}[ht]
\centering
\includegraphics{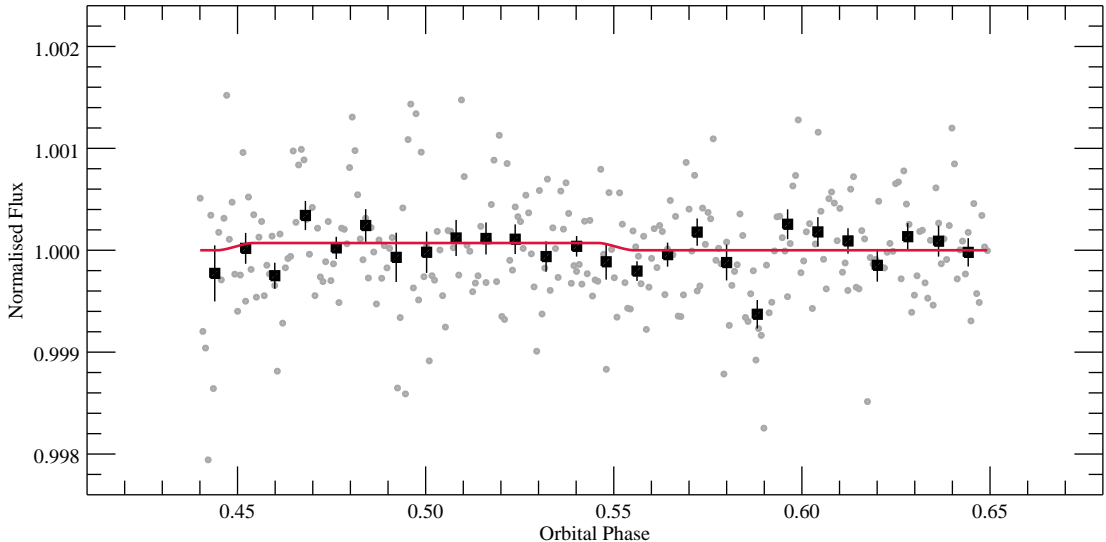}
\caption{The unbinned normalised light curve (grey points) after decorrelating with seeing and a second order polynomial, and the best fit eclipse model (red line). The data binned by phase 0.008 (black squares) are shown for clarity.}\label{Fig2}
\end{figure*}

The raw light curves for the target and each of the comparison stars are shown in the top panel of Figure \ref{Fig1} (a). The two comparisons (HD 195558 and BD+39 4224) are both A-type stars and have median fluxes of 0.27 and 0.12, respectively, relative to KELT-9 throughout the observation. The light curve in the bottom panel of  Figure \ref{Fig1} (a) was created by dividing the target light curve by the sum of the two comparison light curves and normalising for a median value of 1. This step removed visible correlations with airmass and exposure time (top and middle panels of Figure \ref{Fig1} (b)) that were visible in the raw light curves. No correlations were observed with the x and y positions of the stars on the detector. There were small discontinuities visible in the sky background levels for the target (see the bottom panel of Figure \ref{Fig1} (b)) and comparison stars, but no such features were visible in the corresponding light curves. On initial inspection, a dip in the sky background at phase of $\sim$0.59 corresponded with a dip in flux in the normalised light curve (also see Figure \ref{Fig2}). However, the latter feature was found to be much broader and no significant correlation was found. Despite our method of defocusing the telescope, a strong correlation with seeing was visible. There was also a low-order trend present through the time-series that was not removed by the normalisation with the comparison stars.

We fit for each of these trends simultaneously with an eclipse model using a Markov Chain Monte Carlo (MCMC) method with the Metropolis--Hastings algorithm and orthogonal stepping. To model the eclipse, we used the Mandel and Agol transit model \citep{2002ApJ...580L.171M} with the limb darkening coefficients set to zero. The relevant stellar, planetary and orbital parameters were all fixed using values from \citet{2017Natur.546..514G}, which are shown in Table \ref{table}. The trend in the baseline was modelled using a second-order polynomial, which optimised the Bayesian information criterion with respect to higher and lower order models. The time-dependent and time-independent components of the noise associated with the residual flux were measured using the wavelet method \citep{2009ApJ...704...51C}. We ran an MCMC where the components of the model associated with the eclipse depth, a linear function of seeing, the polynomial baseline and the wavelet noise parameters were all allowed to vary. Firstly, we ran a ``burn-in'' phase of 10\textsuperscript{5} steps, where the step sizes were recalculated every 10\textsuperscript{4} steps to set proportionate step sizes for each parameter. We then used these step sizes in an MCMC chain of 10\textsuperscript{6} steps to get the best fit values for each parameter, which are shown in Figure \ref{Fig1} (c). We verified convergence by checking the Gelman--Rubin criterion \citep{1992StaSc...7..457G}.

The detrended light curve is shown in Figure \ref{Fig2}, with the best fit eclipse model shown in red. Whilst a negative value for a secondary eclipse depth is unphysical, we allowed this to avoid biasing the MCMC fit. The best fit depth of -71$\pm$84 ppm allowed us to place an upper limit on the secondary eclipse depth of 181 ppm at 3$\sigma$.

\begin{table}[h]
\begin{centering}
\caption{Parameters of the KELT-9 system} \label{table}
\begin{tabular}{c c c}
\hline
\hline
Parameter & Value & Ref.\\
\hline
\multicolumn{3}{c}{Stellar Parameters} \\
$R_*$ ($R_\odot$) & $2.362^{+0.075}_{-0.063}$ &A\\
$T_{*}$ (K) & 10,170$\pm$450 &A\\
$log(g)$ & 4.091$\pm$0.014 &A\\
$[$Fe/H$]$ & -0.03$\pm$0.20 &A\\
\hline
\multicolumn{3}{c}{Planetary Parameters}\\
$R_P$ ($R_J$) & $1.888^{+0.062}_{-0.052}$ &A\\
$t_0$ (MJD) & 57095.18572$\pm$0.00014 &A\\
$P$ (days) & 1.4811235$\pm$0.0000011 &A\\
$a$ (au) & $0.03462^{+0.0110}_{-0.0093}$ &A\\
$i$ ($^{\circ}$) & 86.79$\pm$0.25 &A\\
$T_{day}$ (K) & 4600$\pm$150 &A\\
$F_{ecl,z'}$ (ppm) & 1006$\pm$97 &B\\
\hline
\multicolumn{3}{c}{Measured Parameters}\\
$F_{ecl,U}$ (ppm) & -71$\pm$84 &C\\
$F_{ecl,U}$ (ppm) & $<$ 181 (3$\sigma$ limit)  &C\\
$T_{day}$ (K) & $<$ 4995 (3$\sigma$ limit)  &C\\
$\sigma_w$ & (3.32$\pm$0.11)x10\textsuperscript{-4}&C\\
$\sigma_r$ & (0.7$\pm$5.6)x10\textsuperscript{-4}&C\\
\hline
\multicolumn{3}{l}{\textbf{References.} A - \citet{2017Natur.546..514G}; B - Collins et al. }\\
\multicolumn{3}{l}{(2019, in prep.); C - This work.}\\
\end{tabular}
\end{centering}
\end{table}

\section{Discussion \& Conclusions}
\label{sec:discussion}

\begin{figure*}[t]
\includegraphics{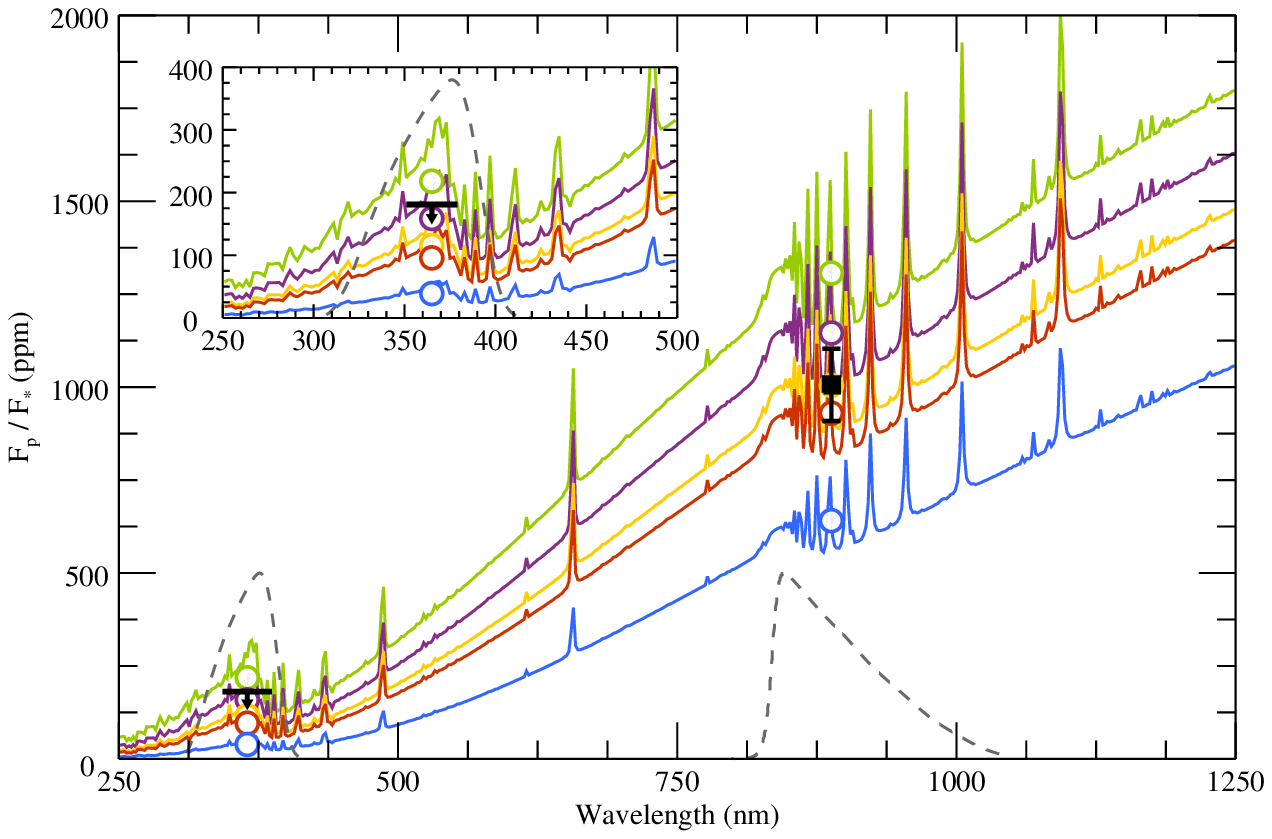}
\includegraphics{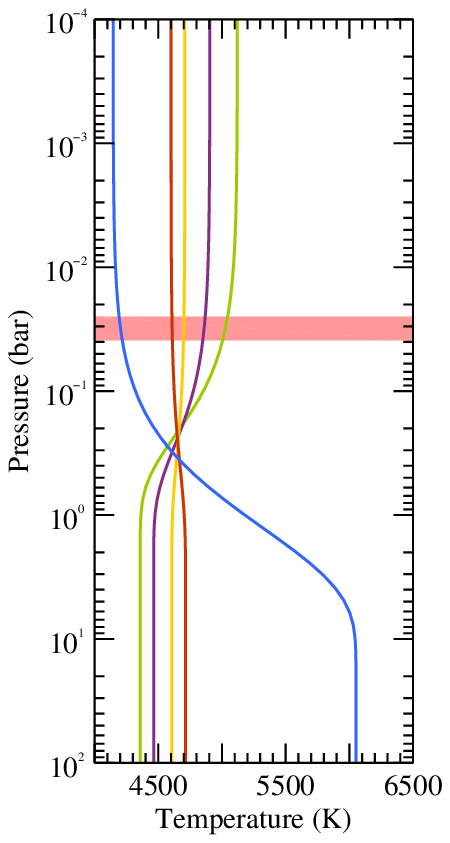}
\caption{\textit{Left}:  Model spectra for the five temperature profiles shown in the right panel, incorporating flux associated with both thermal emission and scattering. Our U-band upper limit and the z'-band eclipse depth (Collins et al. 2019, in prep.) are shown in black and the expected eclipse depths for each profile are marked with coloured data points. The response functions associated with the two bandpasses are marked with grey dashed lines. A close-up of U-band wavelengths is shown inset. \textit{Right}: Temperature-pressure profiles for each of the spectra shown. The red shaded area shows the pressure levels from which the thermal emission for all the models originate, when approximated as blackbodies.}\label{spectra}
\end{figure*}

We jointly interpreted our U-band upper limit and the z'-band eclipse detection (Collins et al. 2019, in prep.) by generating high-resolution emission spectra with wavenumber resolution of 0.03 cm$^{-1}$ (left panel of Figure \ref{spectra}; shown in 2 nm wavelength bins for clarity), which were calculated using a 4-stream discrete ordinate radiative transfer method. For the temperature-pressure (TP) profiles (Figure \ref{spectra}, right panel) we used the approximations for an irradiated atmosphere from \citet{2010A&A...520A..27G}, with an infrared opacity of 0.03 cm$^2$ g$^{-1}$. The ratio of the shortwave to the infrared opacity ($\gamma$), which effectively controls the shape of the TP profile, is treated as a free parameter. Strong absorption of shortwave radiation in the upper atmosphere can result in large temperature inversions of several hundreds of Kelvin, which has previously been demonstrated for species such as TiO and VO \citep{2009ApJ...699.1487S}. Atomic and ionic species expected to be present in the atmosphere of KELT-9b \citep[such as \ce{Fe} and \ce{Fe+};][]{Hoeijmakers:2018ir} are strong absorbers at optical and shorter wavelengths. Hence, the five spectra shown in Figure \ref{spectra} were selected to explore a range of possible TP-profiles, ranging from a strong inversion ($\gamma$=2) to a rapid decrease in temperature with altitude ($\gamma$=0.2). As KELT-9b is tidally locked and expected to inefficiently redistribute heat from dayside to nightside, the temperatures in all five profiles are well above the equilibrium temperature of 4050 K from \citet{2017Natur.546..514G}. The chemical composition is calculated using the \texttt{FastChem} equilibrium chemistry code \citep{2018MNRAS.479..865S}, assuming solar elemental abundances. As shown by \citet{Kitzmann:2018wt}, the assumption of chemical equilibrium is reasonable for the hot dayside of KELT-9b.

We account for about 50 different gaseous absorbers in the atmosphere. Cross-sections for \ce{CO} and \ce{H2O} were calculated with the opacity calculator \texttt{HELIOS-K} \citep{2015ApJ...808..182G}, using the corresponding Exomol line lists. Atoms and ions, including \ce{Fe}, \ce{Fe+}, \ce{Ti}, \ce{Ti+}, \ce{Ca} and \ce{Ca+}, are incorporated with line list data from the Kurucz database. Continuum absorption of \ce{H-} is treated according to \citet{1988A&A...193..189J}. Additionally, we include the collision-induced absorption of \ce{H2}-\ce{H2}, \ce{H2}-\ce{He}, and \ce{H}-\ce{He} pairs, based on data from HITRAN. Furthermore, Rayleigh scattering of \ce{H2}, \ce{H}, \ce{He}, and \ce{CO} is incorporated in the radiative transfer calculations as well.

For KELT-9, we used a spectrum for a 10,000 K star with $log(g)$ = 4 and [Fe/H] = 0 \citep[based on values from][shown in Table \ref{table}]{2017Natur.546..514G} from the NextGen Model grid of theoretical spectra \citep{1999ApJ...512..377H}. 

\begin{table}[h!]
\renewcommand{\thetable}{\arabic{table}}
\centering
\caption{Description of spectra in Figure \ref{spectra}} \label{tab:spectra}
\begin{tabular}{ccccc}
\tablewidth{0pt}

\hline
\hline
Colour & $\gamma$ & $F_{ecl,U}$ (ppm) & $F_{ecl,z'}$ (ppm) & $T_B$ (K)\\
\hline
Green & 2.0 & 218 & 1307 & 5080\\
Purple & 1.5 & 159 & 1146 & 4890\\
Yellow & 1.1 & 116 & 1005 & 4700\\
Red & 0.9 & 96 & 930 & 4600\\
Blue & 0.2 & 39 & 641 & 4165\\ 
\hline
\end{tabular}
\end{table}

We integrated each of the spectra over both filter response functions, which incorporate atmospheric extinction and the quantum efficiency of the detectors, to calculate the expected eclipse depths for each value of $\gamma$ in each band. These are shown on the left panel of Figure \ref{spectra} and listed in Table \ref{tab:spectra}. The effective midpoint of the U-band response function falls at shorter wavelengths than the Balmer jump of KELT-9, boosting the expected eclipse depths in this bandpass. At low resolution for the wavelengths shown, the spectrum associated with each temperature-pressure profile is almost indistinguishable from that of a blackbody spectrum, with the majority of the features visible in the left panel of Figure \ref{spectra} originating in the stellar spectrum. This is due the high atmospheric mixing ratio of \ce{H-} ions: a major source of opacity at NUV, optical and NIR wavelengths \citep{1988A&A...193..189J} that atmospheric models for other ultra-hot Jupiters predict will be present in large quantities \citep[e.g.][]{2018ApJ...855L..30A,2018ApJ...866...27L}. The brightness temperatures for the blackbody spectra most closely matching each of our spectra are also listed in Table \ref{tab:spectra}.

When comparing the brightness temperatures for each spectrum to their corresponding temperature-pressure profiles, it is apparent that each of these temperatures originates at pressure levels of $\sim$30 mbar (right panel of Figure \ref{spectra}; shaded red), independently of the form each of the temperature-pressure profiles takes. Thus, we use our result to place a 3$\sigma$ limit of 4995 K on the dayside temperature of KELT-9b at pressure levels of $\sim$30 mbar, which is in agreement with the value 4600$\pm$150 K in \citet{2017Natur.546..514G}. Whilst two broadband eclipse measurements cannot place strong constraints on the form the temperature-pressure profile of KELT-9b takes at these altitudes, they tentatively favour profiles without a strong temperature inversion. However, the two measurements combined support the theory that the continuum of H- opacity produces an emission spectrum that is very similar to a blackbody at these wavelengths.

When adopting the temperature of 4600$\pm$150 K for the altitudes probed by our observations, we can use our U-band observational data to place a 3$\sigma$ limit of $A_g<$ 0.14. Our models show that the main source of scattering in the NUV is molecular Rayleigh scattering. However, this is expected to be negligible compared to \ce{H-} extinction in the dayside of KELT-9b. \citet{Kitzmann:2018wt} predicted that free electrons are present in significant quantities in the upper atmosphere of KELT-9b, but that the cross-section of Thomson scattering is too small to have a significant contribution to the total flux. When only considering flux from reflected light, integrating over the U-band response function yields predicted eclipse depths of $\sim$3 ppm for all the temperature-pressure structures shown, corresponding to an $A_g$ of $\sim$0.005. Hence, the flux due to reflected light is expected to be negligible compared to the thermal emission in the U-band and well below the detection limits of our observation.

Of the two hot Jupiters for which similar measurements have been carried out, this study suggests that the reflective properties of KELT-9b are more comparable to those of WASP-12b than HD 189733 b. This fits into the wider picture of other hot Jupiters for which a measurement of geometric albedo has been performed, with the vast majority of studies measuring very small components of incident light reflected, which in turn supports studies that suggest that the temperatures present in the daysides of ultra-hot Jupiters are too high for condensates to form. Further secondary eclipse observations in the NUV and blue-optical of HD 189733 b and other hot Jupiters with lower levels of irradiation are required to assess the validity of claims that cloudiness (and therefore observed Rayleigh scattering) scales with temperature in the daysides of hot Jupiters.

Full phase curve observations with space-based facilities such as Spitzer, CHEOPS and TESS will put tighter constraints on the energy budget of KELT-9b and help to break the degeneracy between Bond albedo and atmospheric circulation. Transmission spectroscopy observations at blue-optical wavelengths will test the occurrence of Rayleigh scattering and test how the reflective properties of KELT-9b vary with longitude. Further secondary eclipse observations at optical and IR wavelengths will put tighter constraints on the spectral energy distribution of KELT-9b.

This letter demonstrates the ability of ground-based 2m class telescopes like the INT to perform secondary eclipse observations for hot Jupiters in an age where there will be no spaced-based alternatives, enabling the placement of a tight constraint on the UV eclipse depth from a single observation. For cooler exoplanets than KELT-9b, this will be a direct measurement of their geometric albedos. To date, all previous studies to detect reflected light from transiting exoplanets have been conducted using space-based facilities such as Hubble, Kepler \citep[e.g.][]{2011MNRAS.417L..88K}, CoRoT \citep{2010A&A...513A..76S} and MOST \citep{2008ApJ...689.1345R}. Currently, the only suitable\footnote{The XMM-Newton Optical Monitor \citep{2001A&A...365L..36M} has NUV coverage, but the instrument is designed to supplement simultaneous X-ray observations.} space-based facility with NUV coverage is Hubble, which is likely to permanently go out of operation in the next decade. While operational, although Hubble does not suffer the light extinction in the NUV that ground-based facilities do, the lower time-pressure on ground-based facilities enables easier acquisition of several secondary eclipses, which would yield even higher photometric precision than our single observation.

We thank the anonymous referee for their full and thorough feedback. We are grateful to the staff of the INT for their assistance with these observations. The INT is operated on the island of La Palma by the Isaac Newton Group of Telescopes in the Spanish Observatorio del Roque de los Muchachos of the Instituto de Astrof{\'i}sica de Canarias. MJH acknowledges funding from the Northern Ireland Department for the Economy. CAW acknowledges support from STFC grant ST/P000312/1. NPG acknowledges support from the Royal Society in the form of a University Research Fellowship. DK acknowledges financial and administrative support by the Center for Space and Habitability and the PlanetS National Centre of Competence in Research (NCCR).


\end{document}